\documentclass[aps,twocolumn,showpacs,english,prl,superscriptaddress]{revtex4-1}

\usepackage{graphicx}
\usepackage{dcolumn}
\usepackage{bm}
\usepackage[T1]{fontenc}
\usepackage[latin9]{inputenc}
\usepackage{textcomp}
\usepackage{amsmath}
\usepackage{graphicx}
\usepackage{amssymb}
\usepackage{units}
\usepackage{xcolor}

\begin{document}

\title{Exciton-polaritons gas as a nonequilibrium coolant}

\author{Sebastian Klembt}
\affiliation{Institut N\'{e}el, Universit\'{e} Grenoble Alpes and CNRS, B.P. 166, 38042 Grenoble, France}

\author{Emilien Durupt}
\affiliation{Institut N\'{e}el, Universit\'{e} Grenoble Alpes and CNRS, B.P. 166, 38042 Grenoble, France}

\author{Sanjoy Datta}
\affiliation{LPMMC, Universit\'{e} Grenoble Alpes and CNRS, B.P. 166, 38042 Grenoble, France}

\author{Thorsten Klein}
\affiliation{University of Bremen, P.O. Box 330440, 28334 Bremen, Germany}

\author{Augustin Baas}
\affiliation{Institut N\'{e}el, Universit\'{e} Grenoble Alpes and CNRS, B.P. 166, 38042 Grenoble, France}

\author{Yoan Léger}
\affiliation{Laboratoire FOTON, CNRS, INSA de Rennes, 35708 Rennes, France}

\author{Carsten Kruse}
\affiliation{University of Bremen, P.O. Box 330440, 28334 Bremen, Germany}

\author{Detlef Hommel}
\affiliation{University of Bremen, P.O. Box 330440, 28334 Bremen, Germany}

\author{Anna Minguzzi}
\affiliation{LPMMC, Universit\'{e} Grenoble Alpes and CNRS, B.P. 166, 38042 Grenoble, France}

\author{Maxime Richard}
\affiliation{Institut N\'{e}el, Universit\'{e} Grenoble Alpes and CNRS, B.P. 166, 38042 Grenoble, France}

\pacs{71.36.+c, 78.20.nd, 78.30.Fs}




\begin{abstract}
Using angle-resolved Raman spectroscopy, we show that a resonantly excited ground-state exciton-polariton fluid behaves like a nonequilibrium coolant for its host solid-state semiconductor microcavity. With this optical technique, we obtain a detailed measurement of the thermal fluxes generated by the pumped polaritons. We thus find a maximum cooling power for a cryostat temperature of $50$K and below where optical cooling is usually suppressed, and we identify the participation of an ultrafast cooling mechanism. We also show that the nonequilibrium character of polaritons constitutes an unexpected resource: each scattering event can remove more heat from the solid than would be normally allowed using a thermal fluid with normal internal equilibration.
\end{abstract}

\date{\today}

\maketitle

Owing to the selection rules in light-matter interaction, light can pick-up energy and momentum from matter and conversely, as soon as translational invariance is broken by e.g. a dielectric interface or a point-like dipole like an atom. This seemingly trivial property is at the basis of several spectacular achievements such as Doppler cooling of atomic gases \cite{Wieman}, optical tweezers \cite{Keir}, or cavity mirror motion cooling by radiation pressure \cite{Gigan,Arcizet,Chan}. Long before such ideas were implemented, Peter Pringsheim proposed in 1929 that such an effect could be used to cool down solids \cite{Pringsheim}. Indeed, discrete translational invariance in ideal solids is broken by thermal vibrations of the lattice (phonons), the energy of which can be transferred to light. Such a mechanism is known nowadays as Anti-stokes fluorescence (ASF): a laser is tuned to the vibrational ground state of an electronic transition, such that the re-emitted photons are likely to take away with them the energy of one or several thermal phonons, resulting in a temperature drop of the material. The technologic advantages of such a technique are obvious, like the absence of moving mechanical parts, and the very convenient nature of light as a coolant, easy to create, dissipate, and propagate across large distances.

\begin{figure}[t]
\includegraphics[width=\columnwidth]{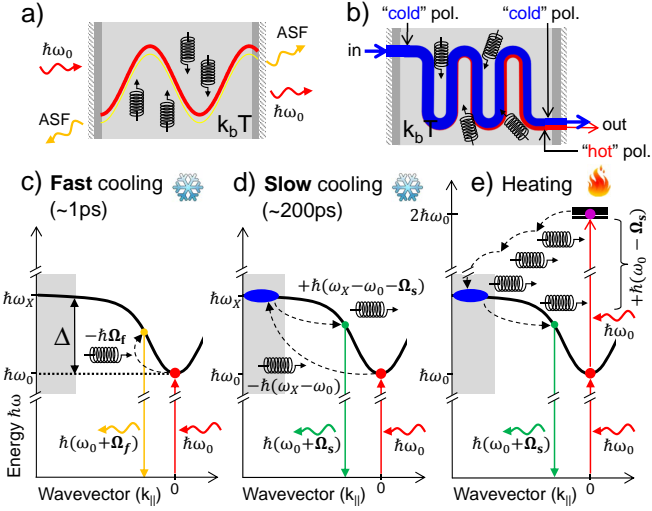}
\caption{(Color online) a) Principle of the cooling method: the polariton field absorbs thermal phonons resulting in polariton ASF emission. The wavy arrows depict photons, while the spring-like arrows represent phonons. b) Thermodynamical description: during its lifetime, a small fraction of the "cold" polariton gas is "heated up" by phonons into a non-thermal state. c), d) and e) summarize the three heat exchange mechanisms set in motion by the cooling method. Phonon energy is counted positive (negative) for emission (absorption). \emph{Fast} cooling mechanism b) removes phonons of average energy $\hbar\Omega_f$, \emph{Slow} cooling mechanism c) removes phonons of average energy $\hbar\Omega_s$, while two-photon absorption d) generates a phonon cascade (heat) of average energy $\hbar\omega_0-\hbar\Omega_s$. The gray rectangle is the area outside the polariton light cone.}
\label{fig1}
\end{figure}

In order to implement this strategy, the material must feature an electronic transition with a high radiative rate and quantum efficiency (i.e. a vanishing non-radiative recombination). Owing to their imperfections, these conditions are hard to meet in real materials. The best results so far have been obtained in Ytterbium-doped crystals. Indeed, in spite of the long radiative lifetime and weak oscillator strength of the embedded Ytterbium atoms, their high quantum efficiency allowed cooling from room temperature down to $T=110$K \cite{Sheik1,110K}. In semiconductor materials, the excitonic transition has a much larger oscillator strength, a stronger coupling to phonons, and a shorter radiative rate. It was thus predicted that optical cooling in semiconductors should be more efficient \cite{Rupper,Sheik2}. However, despite encouraging results \cite{Gauck,Finkeissen} it is only recently that a room temperature ZnS nanostructure could be cooled down to $T=260K$ \cite{Zhang}. The difficulty lies in the fact that, owing to their extended nature, excitons are more prone to multiphonon non-radiative relaxation via impurity or defect states. These relaxation channels involve phonon cascade emission \cite{nr1} that competes with the phonon absorption achieved by ASF, making it hard to reach a positive net cooling power.

In this work, instead of bare photons, we consider using resonant optical pumping of ground state exciton-polaritons to excite polariton ASF, and thus generate a cooling power in a solid-state semiconductor microcavity (MC) in the strong coupling regime (cf. Fig.\ref{fig1}.a). Polaritons are bi-dimensional quasi-particles that benefit from a half-photonic, half-excitonic nature \cite{weisbuch}, to achieve much stronger interaction with phonons than photons \cite{fainstein,jusserand}. Fig.\ref{fig1}.c shows the principle of the most interesting cooling mechanism achieved by this excitation scheme, and that we report in this letter: within a short lifetime of $\sim 1$ ps, a polariton can be excited into a higher energy state by absorbing a thermal acoustic phonon of average energy $\hbar\Omega_f$. Its subsequent radiative recombination results in the net energy transfer of $\hbar\Omega_f$ from the thermal phonon bath to the outside electromagnetic vacuum. Two additional thermal exchange mechanisms also take place, a cooling one shown in $Fig.\ref{fig1}.d$, and a heating one shown in $Fig.\ref{fig1}.e$, as will be discussed later.

Polaritons have many advantages over other optical methods to cool down semiconductor materials: their cavity-like dispersion (cf. $\omega(k_\parallel)$ in Figs.\ref{fig1}.c-e) fully inhibits the Stokes emission which is a source of phonon emission. Moreover, owing to their bi-dimensional degree of freedom, a 2D continuum of states is available for anti-Stokes scattering, so that thermal phonons of arbitrary low energy can be removed in the process, thus preventing a cooling power cutoff at low temperatures. Another useful property of polaritons in this context is their ultra-light mass of $\sim 10^{-4}$ in electron mass unit: it allows anti-Stokes polaritons to remain within the light cone even when high energy thermal phonons are involved (cf. Fig.\ref{fig1}.c), and it quenches the coupling with point-like defects that could cause non-radiative recombination. We verified indeed that this relaxation channel could be safely neglected in this thermal analysis \cite{supp_mat}. Finally, the strong coupling regime reduces the scattering of polaritons towards the long-lived dark exciton level \cite{darkX} that favors non-radiative recombination. In this context, a large Rabi splitting is desirable. We thus fabricated a high-quality Selenide-based MC for our experiment similar to that used in ref. \cite{sebald}, displaying a Rabi splitting of $\hbar\Omega_R=29$meV, stable for any cryostat temperature ranging from $T=5$K to $T=150$K. In this experiment, polariton cooling is set in motion by shining a linearly polarized CW laser beam at normal incidence on the MC, focused in a spatially Fourier-transform spot of $20\mu$m diameter, and resonant with $k_\parallel=\sqrt{k_x^2+k_y^2}=0$ polaritons at $\hbar\omega_0=2791.1meV$.

Our aim is not to directly measure a temperature drop of the MC. Such a task is indeed technologically challenging: one first needs to grow a microcavity with a non-absorbing substrate on the back side, or needs to remove it. Secondly, suspended microstructures must then be designed and etched in order to achieve the best possible thermal insulation from the rest of the sample. Secondly, suspended microstructures must then be designed and etched in order to achieve the best possible thermal insulation from the rest of the sample. In this work, in order to demonstrate the principle and to characterize the performance of this cooling method, we use polariton ASF to perform a detailed measurement of the different thermal fluxes generated within the MC by the pumped polaritons. Since polariton ASF is isotropic and depolarized by anti Stokes scattering \cite{supp_mat}, it is measured by angle-resolved, cross-polarized spectroscopy in reflection configuration. We choose a negative cavity -exciton energy detuning $\delta=-17meV$ (at $T=4.2$K) of the MC so that the polariton dispersion spans over an energy range of $\Delta=\hbar\omega_X-\hbar\omega_0=26$meV (cf. Fig.\ref{fig1}.c). In this way, at the temperature we are interested in, most of the thermal phonons have an energy lower than $\Delta$ and thus create anti-Stokes polaritons within the light cone. The small remaining fraction creates bare excitons at large momenta as shown in Fig.\ref{fig1}.d. These two mechanisms will be labeled as \emph{fast} and \emph{slow} cooling respectively thereafter.

\begin{figure}[t]
\includegraphics[width=\columnwidth]{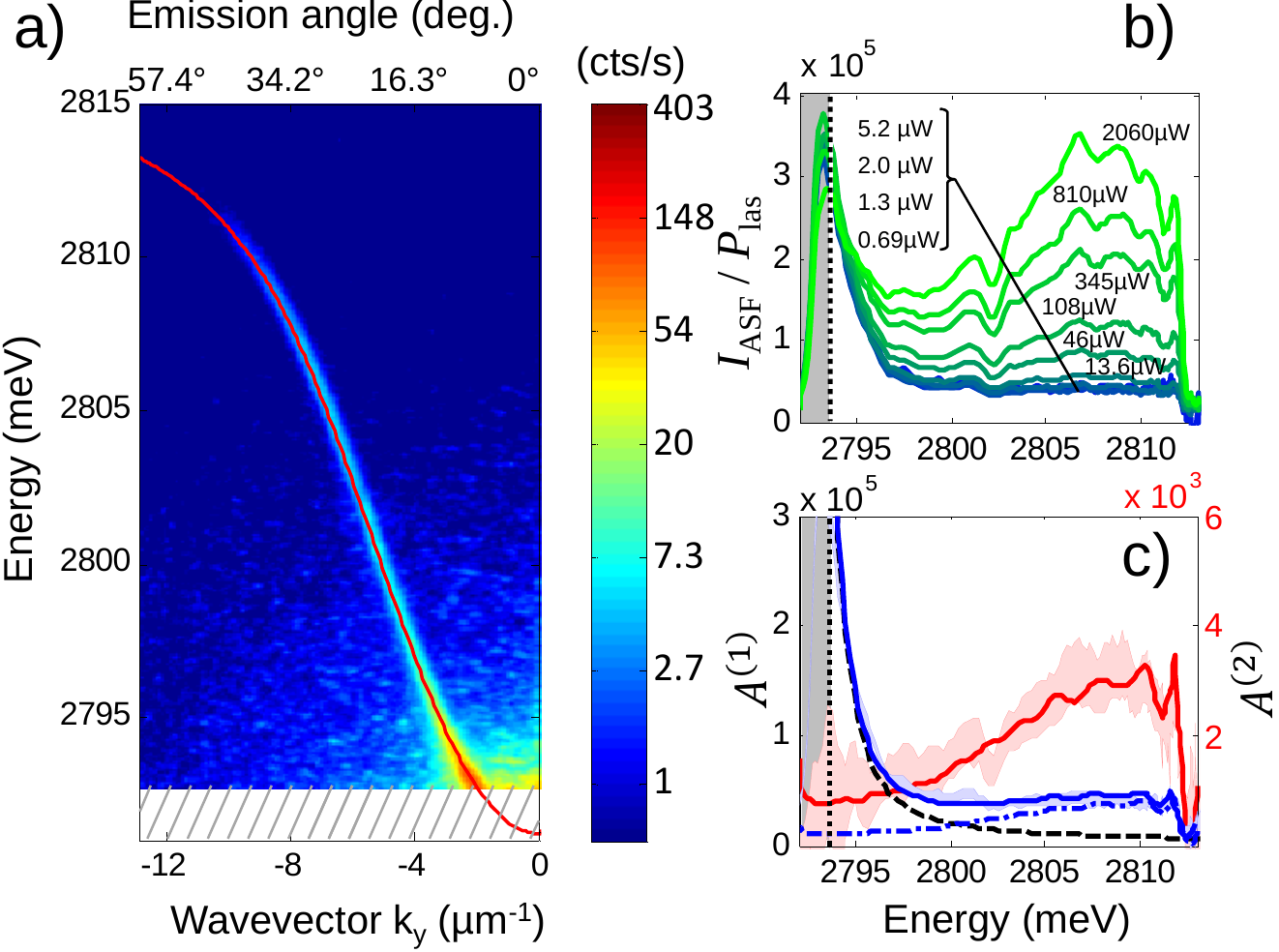}
\caption{(Color online) a) ASF spectrum at $T=20$K, in logarithmic color scale versus $k_y$ and $\hbar\omega$. the corresponding emission angle $\theta$ is shown on the top axis. The red line is a fit to the dispersion shape. b) ASF spectral density divided by $P_\text{las}$ in cts/s/meV/$\mu$W. Solid lines color evolves from blue to green for increasing leaser power $P_\text{las}$. c) Measured (blue stripe) and fitted (solid blue line) linear component $A^{(1)}(\omega)$ of the spectrum in cts/s/meV/$\mu$W. Red: measured (red stripe) and fitted (solid red line) quadratic component of the spectrum in cts/s/meV/$\mu$W$^2$. The solid red line is proportional to $I_\text{PL}(\omega)$. The dashed black line is the \emph{fast} cooling component in $A^{(1)}(\omega)$, while the dash dotted blue line is the \emph{slow} one.}
\label{fig2}
\end{figure}

A typical raw ASF spectrum is shown in Fig.\ref{fig2}.a for a cryostat temperature of T$=20$K in logarithmic color scale. It is obtained by spectrally resolving a narrow slice (along $k_y$) of the ASF emission, where residual laser light is further rejected by spectral and $k$-space filtering (cross-hatched region at the bottom of Fig.\ref{fig2}.a. A careful calibration of the whole setup optical transmission allows us to obtain the ASF spectrum $i_\text{ASF}(\omega)$ in absolute anti-Stokes photon emission rate in counts per seconds. Finally, since ASF emission is isotropic, the total ASF spectral density $I_\text{ASF}(\omega)$ within a cone of $8.4^\circ \leq \theta\leq 43^\circ$ radius is extrapolated. $I_\text{ASF}(\omega)$ normalized by the excitation power $P_\text{las}$ is plotted in Fig.\ref{fig2}.b for different $P_\text{las}$. Details on the experimental setup and data treatment are given in the supplemental materials \cite{supp_mat}.


A first striking behavior is that the high energy part of the spectra behaves nonlinearly with respect to $P_\text{las}$. A careful numerical analysis of this dataset allows to extract $A^{(1)}(\omega)$ and $A^{(2)}(\omega)$, i.e. the fraction of the ASF spectrum that behaves linearly and quadratically respectively with respect to $P_{las}$ \cite{supp_mat}. As shown in Fig.\ref{fig2}.c, $A^{(2)}(\omega)$ has the same spectral shape as a photoluminescence spectrum $I_\text{PL}(\omega)$ (solid red line in Fig.\ref{fig2}.c) obtained under non-resonant excitation with a CW laser tuned at $\hbar\omega_\text{nr}=2990$meV, i.e. above the band to band transition. $A^{(2)}(\omega)$ thus clearly results from two-photon absorption of the pump laser under the form of high energy free carriers, followed by a phonon cascade that relaxes the carriers back into the polariton states (i.e. the mechanism shown in Fig.\ref{fig1}.e). From the thermal balance point of view, this mechanism is a source of heat since each absorption event adds a cascade of phonons of total average energy $\hbar(\omega_0-\Omega_s)$ into the lattice, where $\hbar(\omega_0+\Omega_s)$ is the average energy of $I_\text{PL}(\omega)$. Fortunately, its quadratic behavior with respect to $P_\text{las}$ makes its contribution negligible at low enough excitation power $P_\text{las}$.

$A^{(1)}(\omega)$, the part of the spectrum which is linear with $P_{las}$, results from the absorption of thermal phonons by ground state polaritons of energy $\hbar\omega_0$ followed by fluorescence at higher energy. It thus provides the wanted cooling power, and involves two distinct mechanisms. In the first one, that we call \emph{fast}, a ground state polariton is scattered directly into an excited polariton of average energy $\hbar(\omega_0+\Omega_f)$, that recombines radiatively within a timescale of $\sim 1$ ps (cf. Fig.\ref{fig1}.c). The second one, that we call \emph{slow}, has the excitonic reservoir (outside the light cone) as a longer lifetime ($\sim 200$ ps \cite{langbein}) intermediate state (Fig.\ref{fig1}.d). Within a spectral analysis of $A^{(1)}(\omega)$, the proportion of these two different contributions can be evaluated quantitatively. Indeed, a Fermi golden rule approach provides an accurate simulation of the \emph{fast} ASF spectrum $I_\text{th}(\omega)$ (dashed line in Fig.\ref{fig2}.c) that shows that its contribution is maximum at low energy (only the high energy flank is visible experimentally). On the other hand, the \emph{slow} ASF spectrum is well accounted for by the experimentally measured non-resonant spectrum $I_\text{PL}(\omega)$, which exhibits a pronounced bottleneck $\sim 15$ meV above the ground state (dash-dotted blue line in Fig.\ref{fig2}.c). We clearly see in $A^{(1)}(\omega)$ these two contributions: the upper flank of a peak on the low energy side energy (\emph{fast} ASF) and a shoulder on the high energy side (\emph{slow} ASF). Using both $I_\text{th}(\omega)$ and $I_\text{PL}(\omega)$ to fit $A^{(1)}(\omega)$ (cf. blue solid line in Fig.\ref{fig2}.c), the fraction of \emph{fast} cooling $\rho$ contributing to the overall cooling power is thus evaluated. The accuracy of this procedure is assured by the fact that $I_\text{th}(\omega)$ and $I_\text{PL}(\omega)$ have a quite different spectral shape. Note that since the microcavity parameters, $I_\text{PL}(\omega)$, as well as the materials parameters are measured or known from the literature (cf. Supplemental information) only the overall multiplicative factors of $A^{(1)}(\omega)$ and $A^{(2)}(\omega)$ are free fitting parameters.

\begin{figure}[t]
\includegraphics[width=\columnwidth]{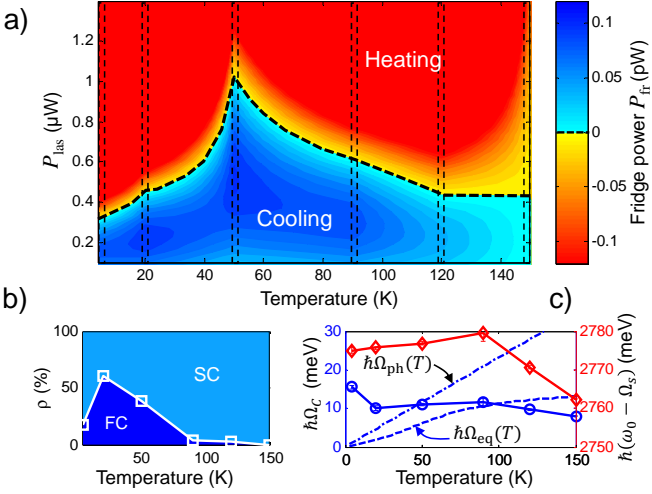}
\caption{ (Color online) a) Total cooling power $P_\text{fr}$ versus cryostat temperature (T) and laser power $P_\text{las}$. The dashed vertical rectangles contains color plots of the measured function $P_\text{fr}(P_\text{las})$ (see main text and eq.(1) for the method). The rest of the color plot is an interpolation serving as a guide to the eye. The dashed line separate the cooling region from the heating one. b) participation $\rho$ of \emph{fast} cooling (FC) (SC stands for \emph{slow} cooling) involved in the overall cooling process. c) Red crosses: average energy $ \hbar\omega_0-\hbar\Omega_s$ of the phonon cascade involved in the two-photon absorption heating. Blue hollow circles: average phonon energy $\hbar\Omega_c$ removed by the overall cooling process. The dashed line shows the theoretical average energy of polaritons $\hbar\Omega_\text{eq}(T)$ assuming thermal equilibrium with the phonon bath (same temperature $T$). $\hbar\Omega_\text{ph}(T)$ is the average energy of thermal phonons.}
\label{fig3}
\end{figure}

We have carried out this analysis at different cryostat temperatures. We found that the \emph{fast} cooling mechanism has a peak contribution of $\rho=61\%$ at $T=20$K, and remains significant up to $T\simeq 100$ K (cf. Fig.\ref{fig3}.b). This \emph{fast} cooling mechanism is unique to polaritons: its timescale is fixed by the polariton lifetime which is two orders of magnitude shorter than the \emph{slow} cooling one. This dynamics is even faster than the phonons typical thermalization time \cite{phonon_tau}. The observation of this \emph{fast} cooling mechanism in a MC in the strong coupling regime is the key result of this letter. The \emph{slow} cooling mechanism is reminiscent from that involved in exciton-enhanced optical cooling \cite{Rupper}. At such low temperatures, its contribution can seem surprising as it involves thermal phonons of energy comparable with $\Delta \gg k_bT$. However, the weak phonon population at this energy is compensated by two aspects: (i) the excitonic density of states, which is four orders of magnitude larger than that of polaritons, and (ii) the fact that for such large $\Delta$'s the phonons involved are of the optical type, and thus exhibit a coupling strength with polaritons fifty times larger than the acoustic ones involved in the \emph{fast} cooling \cite{rudin}.

Now that we have a measurement of the three main mechanisms involved in the heat exchange between polariton and thermal phonons, we can derive an experimental value of the thermal energy removed from the MC per units of time (i.e. the cooling power) as
\begin{align}
P_\text{fr}(P_\text{las},T)=\int \mathrm{d\omega\,} & \big\{ P_\text{las}\hbar(\omega-\omega_0)A^{(1)}(\omega,T)\nonumber\\
& -P^2_\text{las}\hbar(2\omega_0-\omega)A^{(2)}(\omega,T) \big\},
\end{align}
where $\hbar(\omega-\omega_0)$ is the energy of a phonon which has been removed when a photon is detected at the energy $\hbar\omega$. Positive $P_\text{fr}$ means cooling while negative means heating. Note that in this approach, $P_\text{fr}$ is a lower bound of the true cooling power since due to the objective finite numerical aperture, only $46\%$ of the whole momentum space is accessible. Indeed, the inaccessible large $k_\parallel$ ring involves the absorption of high energy thermal phonons, that have a large contribution to cooling via the \emph{slow} channel. We extracted $P_{\text{fr}}$ for temperatures ranging from $T=4.2$K to $T=150$K. The result is summarized in Fig.\ref{fig3}.a versus temperature and laser power. We find that the maximum cooling power $P_\text{fr}^\text{max}=(0.10\pm0.02)pW$, with $\rho=40\%$ is achieved at $T=50$K, and remains positive below. As explained above, this is only possible because polariton ASF does not involve a discontinuous electronic density of states but rather a continuous one with no energy gap between the pump polaritons and the available anti-Stoke states. Therefore, phonons of very low energy can be pumped out from the thermal bath by this method. Such a behavior is another unique feature of polariton cooling and constitutes the second key result of this letter. Note that this cooling power is generated within a micron-scale volume resulting in a large cooling power density of $p_\text{fr}=(80\pm 16)\mu$W.cm$^{-3}$. Finally, we see in Fig.\ref{fig3}.c (red diamond) why the main limitation to polariton cooling is two-photon absorption, indeed although this absorption rate is low as compared to the cooling rate (at least at power $P_\text{las}$ low enough), each such event releases a large amount of heat $\hbar(\omega_0-\Omega_s)$ in average, ranging from $2760$ meV to $2780$ meV.


According to these results and our understanding of polariton cooling, both the \emph{fast} cooling participation ratio $\rho$ and the cooling power $P_\text{fr}$ could be largely increased by making two fairly simple changes in the MC design. Firstly, by lowering the quality factor $Q=\omega_0/(2\pi\gamma)$ (presently, $Q=5600$), the ratio of two-photon absorption rate over cooling rate would decrease since the earlier scales like $1/\gamma^2$ and the latter scales like $1/\gamma$. Obviously, $Q$ cannot be decreased to arbitrarily low values as it must remain large enough for the strong coupling regime to be preserved. Secondly, when $\Delta$ exceeds the energy $\hbar\Omega_\text{LO}^{max}$ of the highest frequency optical phonon mode, the scattering of a polariton into the excitonic reservoir by a single phonon becomes forbidden by energy conservation. A strong suppression of the \emph{slow} cooling mechanism in favor of the \emph{fast} one is thus achieved. Such a suppression has been demonstrated already in a different context in a ZnO MC \cite{trichet}. An additional condition required to preserve the \emph{fast} cooling mechanism is that the excitonic fraction should remain significant. Both conditions are easily met in state-of-the-art Selenide and Telluride microcavities that combine large Rabi splitting and low LO phonon energy.

This work also gives us a striking insight on the thermal properties of a nonequilibrium cryogenic fluid (polaritons) interacting with a thermal bath (phonons) over a timescale too short for it to thermalize. The thermodynamical point-of-view on this experiment is summarized in Fig.\ref{fig1}.b. Polaritons are injected with an effective temperature much colder than that of thermal phonons (the pump laser injects polaritons in their ground state). During the cooling mechanisms, polaritons pick up heat from the thermal phonon bath, and then recombine radiatively. The steady-state polariton gas resulting from this interaction is highly non-thermal: it consists of two independent components: a "cold" gas that did not interact with phonons and recombines at the same energy at which it entered the MC, and a smaller "hot" one, in which the captured heat is distributed according to a non-thermal distribution function, a measurement of which is given by $A^{(1)}(\omega)$. This "hot" and "cold" components do not mix up like in a normal fluid, because at the low densities involved in this cooling scheme, polaritons do not interact with each others. The properties of this "hot" component are rather unusual: in Fig.\ref{fig3}.c we plotted $\hbar\Omega_c(T)$ the average energy of $A^{(1)}(\omega)$, that represents the average energy removed from the solid per scattering event. $\hbar\Omega_\text{eq}(T)$, the average energy of an hypothetical polariton gas at thermal equilibrium is shown on the same plot. We find that between $T=4.2$K and $T=100$K $\hbar\Omega_c>\hbar\Omega_\text{eq}$, meaning that the thermal energy removed from the solid by each polariton of the "hot" component is larger than if they were at thermal equilibrium with the lattice. We can also compare $\hbar\Omega_c(T)$ with $\hbar\Omega_\text{ph}(T)\simeq 2.701k_bT$, the average energy of the thermal phonon bath. It appears that between $T=4.2$K and $T=50$K, $\hbar\Omega_c>\hbar\Omega_\text{ph}$, meaning this time that the "hot" component of the polariton gas is even "hotter" than the phonon bath itself. This surprising result suggests that introducing a nonequilibrium character to a coolant fluid might constitute a useful resource to enhance its performance.

In this work we have shown, as an experimental proof-of-principle, that polaritonic cooling of a semiconductor MC works and presents unusual properties with respect to state-of-the-art methods of optical cooling in solids. Firstly, owing to the specific polaritonic density of states, that allows removing low energy thermal acoustic phonons, cooling can be achieved at arbitrarily low cryostat temperature. Secondly, polaritonic cooling involves a new mechanism referred to as \emph{fast} cooling that is unique to polaritons, and opens up an experimental window on nonequilibrium thermodynamics of phonons. Finally, we have shown that polariton fluids constitute an experimental model system to study the heat transport properties between a thermal bath and a nonequilibrium fluid.

ED and SK have contributed equally to this work. All the authors acknowledge support by the ERC StG contract Nb 258608. MR, AM, SD, SK, AB, YL and ED wish to thank M. Wouters, I. Carusotto, L. S. Dang, J. Kasprzak, J. Bloch and A. Amo for inspiring discussions. Technical support by C. Bouchard and L. Del-Rey is warmly acknowledged.

\appendix*
\begin{widetext}

\section{SUPPLEMENTAL MATERIAL FOR: Exciton-polaritons gas as a nonequilibrium coolant}

\subsection{Experimental setup and Numerical data treatment}

The raw polariton ASF data are acquired and treated in the following way.
\begin{itemize}

\item
The sample is placed in vacuum inside a variable temperature cryostat used to set the MC base temperature. it is mounted on its holder with an upward angle of $13^\circ$ with respect to the excitation/detection objective optical axis. In this way, spurious reflection of the laser on the objective lenses are suppressed upon exciting at $k_\parallel=0$, and more importantly, ASF detection is possible up to an emission angle of $43^\circ$ (the microscope objective has a numerical aperture of 0.5). The CW laser beam is injected at a well defined incidence angle on the MC by focusing it with a $f=300$mm lens onto the input Fourier plane of the objective, so that a spatially Fourier transform excitation spot is achieved, with a diameter of $20\mu$m on the surface of the sample and an angular spread of $3^\circ$. The laser beam is tuned, angle-wise and in wavelength at resonance with the polariton ground state $(k_\parallel=0,\omega_0)$, where $\hbar\omega_0=2790$ meV.

\item
The excitation laser and the detection are cross-polarized by Glan-Thompson polarizers, thus achieving a rejection efficiency of $10^{-7}$. However, while the laser is fully suppressed with respect to the ASF intensity, a bright cross-polarized polariton emission occurs from the ground state, likely due to a weak local spin-anisotropic disorder. This resonant photoluminescence is three orders of magnitude weaker than the incoming laser but still three to four orders of magnitude brighter than the ASF. To further reject this signal, we use two consecutive filtering steps: the emission Fourier plane is imaged on the entrance slit of the monochromator. It is shifted by $80\mu$m with respect to $k_\parallel=0$ such that the emission at $k_\parallel \leq 0.58\mu$m$^{-1}$ is blocked. The remaining spurious resonant signal entering the monochromator is then shifted away spectrally by $1.65$meV with respect to the bottom of the polariton dispersion, from the edge of the charged coupled device (CCD) sensitive area. Overall, these filtering steps reject emission angles between $0^\circ$ and $8.4^\circ$, while ASF is measured from $8^\circ$ to $43^\circ$.


\item
Absolute calibration of the setup transmission is carried out in the following way: a calibrated silicon photodetector is placed on the transmitted port of the last beamsplitter, while the reflected port leads to the microscope objective. We thus obtain an accurate reading of the laser power $P_\text{1}$ at this point of the setup. The power actually entering the MC is $P_\text{las}=P_\text{1}T_\text{obj}T_\text{pol}$, where $T_\text{obj}=0.75$ is the objective transmission at $\hbar\omega_0$ and $T_\text{pol} \simeq 1-R_\text{pol} $ is the measured MC transmission at resonance. Then the optical efficiency of the setup is calibrated using a reflection of the laser on the MC at a wavelength redshifted from  the polariton resonance where it behaves as a $\sim 100\%$ mirror. Using again our calibrated photodetector, when a photon is emitted by the MC within the detected emission cone, it has a probability of $1.11\%$ to be detected on the CCD. In other terms, at this wavelength, a conversion efficiency of $2.48\times10^{10}$ counts/s on the CCD is found per $\mu$W of fluorescence.

\item
The image thus obtained on the CCD is a measurement of polariton ASF with spectral (Horizontal axis of the CCD) and angular resolution (vertical axis). The function converting the vertical pixels into true wavevector $(k_x,k_y)$ is known by a previous CCD-pixel-to-angle calibration of the microscope objective. The CCD image is thus reshaped according to this function and results in image like that of Fig.2.a (main text). In order to get the polariton ASF count rate $I_\text{ASF}$, the lower polariton branch is fitted with the theoretical one (obtained from the usual two coupled harmonic oscillators model). From that fit, a mask is created that borders the raw dispersion, and rejects the surrounding noise. $i_\text{ASF}(\omega)$, the ASF count rate through the monochromator slit, is then obtained by summing up the counts along $k_y$. The total emission rate $I_\text{ASF}(\omega)$ (cf. Fig.2.b of the main text) is then extrapolated assuming isotropic emission (cf. section 2 of this document).

\item
For a given temperature, the fractions of the ASF spectrum that behave linearly $A^{(1)}(\omega)$ and quadratically $A^{(2)}(\omega)$ with respect to $P_\text{las}$ are obtained from the dataset $I_\text{ASF}(\omega,P_\text{las})$: for each energy pixel $\omega_n$ the data points $I_\text{ASF}(\omega_n,P_\text{las})$ are fitted with the function $A^{(1)}(\omega_n)P_\text{las}+A^{(2)}(\omega_n)P_\text{las}^2$, where $A^{(2)}(\omega_n)$ and $A^{(2)}(\omega_n)$ are the fitting parameters. The thus obtained spectral densities $A^{(1)}(\omega)$ and $A^{(2)}(\omega)$ characterizes the ASF response to optical excitation in counts/s/meV$\mu$W and counts/s/meV$\mu$W$^2$ respectively (cf. Fig.2.c of the main text). The error bars are obtained from the $95\%$ confidence bound of this fitting procedure. This method allows an accurate separation of the two-photon absorption related source of heat, from the source of cooling, namely the \emph{fast} and \emph{slow} cooling mechanisms depicted in Fig.1.a and Fig.1.b of the main text.

\item
Finally to separate the \emph{fast} from the \emph{slow} cooling contribution in the linear part of the ASF spectrum $A^{(1)}(\omega)$, we take advantage of the fact that both mechanisms lead to very different spectra, in particular at this detuning where the non resonant polariton emission (as expected for the \emph{slow} cooling fraction of the ASF) presents a strong bottleneck, i.e. an intensity maximum high above the ground state ($>15meV$ in our case). Indeed the \emph{fast} cooling fraction of the ASF is on the contrary peaked very close to the ground state (at about $0.8meV$) and then decays quasi-exponentially with energy. Thus in order to extract the fraction $\rho$ of \emph{fast} cooling in $A^{(1)}(\omega)$, the latter is fitted with the weighted sum of the non-resonant photoluminescence spectrum $I_\text{PL}(\omega)$ obtained by exciting the MC with a CW non-resonant laser at $2990meV$, and the ASF theoretical spectrum due to the \emph{fast} cooling mechanism $I_\text{th}(\omega)$. In other words, $A^{(1)}(\omega)=C\rho I_\text{th}(\omega)+C(1-\rho)I_\text{PL}(\omega)$, where $C$ is a constant.

\item
The average phonon energy absorbed by the \emph{fast} cooling process reads
\begin{equation}
\hbar\Omega_f=\frac{\sum_n \hbar(\omega_n-\omega_0)I_\text{th}(\omega_n)}{\sum_n I_\text{th}(\omega_n)}
\end{equation}

The average phonon energy absorbed by the \emph{slow} cooling process reads
\begin{equation}
\hbar\Omega_s=\frac{\sum_n \hbar(\omega_n-\omega_0)I_\text{PL}(\omega_n)}{\sum_n I_\text{PL}(\omega_n)}
\end{equation}

The average phonon energy absorbed by both cooling processes reads
\begin{equation}
\hbar\Omega_c=\frac{\sum_n \hbar(\omega_n-\omega_0)A^{(1)}(\omega_n)}{\sum_n A^{(1)}(\omega_n)}
\end{equation}

The average phonon cascade energy emitted after a two-photon absorption process reads
\begin{equation}
\hbar\Omega_2=\frac{\sum_n \hbar(2\omega_0-\omega_n)A^{(2)}(\omega_n)}{\sum_n A^{(2)}(\omega_n)}
\end{equation}
and since $A^{(2)}(\omega)$ has the same shape as $I_\text{PL}(\omega)$, $\hbar\Omega_2=\hbar(\omega_0-\Omega_s)$.

\end{itemize}

\subsection{Polarization properties of the polariton anti-Stokes fluorescence}

\begin{figure}[t]
\includegraphics[width=17cm]{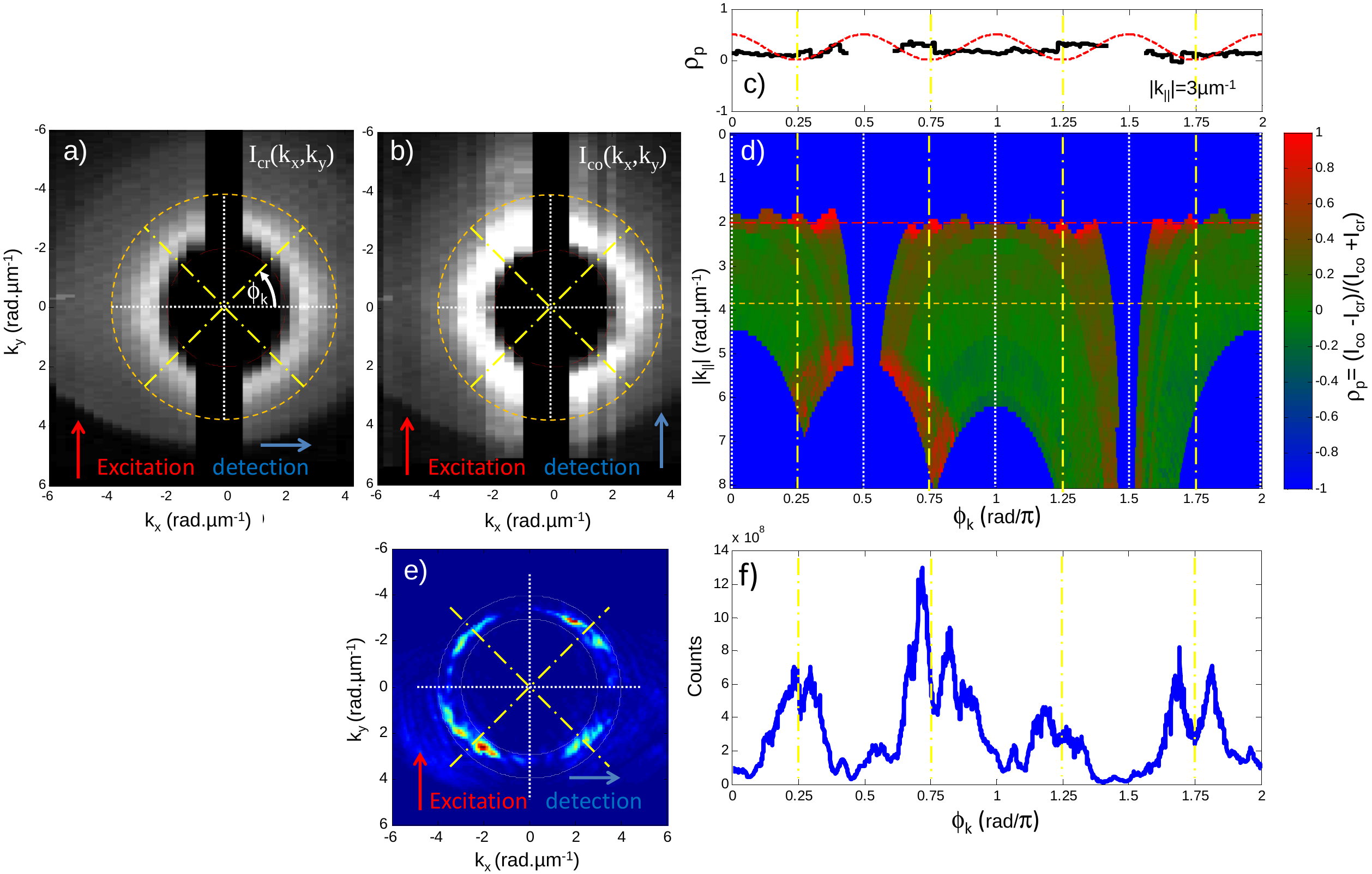}
\caption{(Color online) Spectrally-integrated cross-polarized $I_{cr}(k_x,k_y)$ a) and co-polarized $I_{co}(k_x,k_y)$ b) ASF emission at $T=50$K in $k_\parallel$-space. The hole in the middle (inside the red dashed line) is caused by the method for laser filtering. Inside the area encircled by the red and orange dashed line, \emph{fast} ASF constitutes at least $50\%$ of the counts. c) black solid line, measured degree of linear polarization $\rho_\text{P}(\phi_k)$ for $k_\parallel=3\mu$m$^{-1}$. The red dashed line shows the theoretical linear polarization degree (assuming $50\%$ of ASF) expected assuming only the effect of the TE/TM splitting on the anti-Stokes scattering. d) measured degree of linear polarization $\rho_\text{P}(\phi_k,k_\parallel)$. Green means depolarized. The solid blue area corresponds to points beyond the detection area. Its wavy irregular boundary is due to the sample tilt $\theta_t=13^\circ$ with respect to the objective optical axis. e) cross-polarized resonant fluorescence intensity, for laser excitation at $k_\parallel=3\mu$m$^{-1}$. The corresponding integrated intensity $I_\text{R}(\phi_k)$ of the fluorescence inside the white lines bounded ring is plotted in f). }
\label{figS2}
\end{figure}


Our method of rejection of the laser to detect the weak ASF signal relies on a cross-polarized detection scheme. For the \emph{slow} cooling mechanism and the two-photon absorption related fluorescence, a depolarized emission is expected since reservoir excitons and a fortiori hot free carriers undergo a fast spin scrambling. Therefore a cross-polarized detection scheme allows measuring one half of the counts of such origins. In the \emph{fast} cooling mechanism, a pump polariton is scattered into an excited polariton by absorption of a single thermal phonon. In principle, in this inelastic scattering process, polariton spin flip is strongly suppressed as compared to bare excitons. The reason is the following: acoustic phonons are lattice deformations that do not interact with the carriers spin. They do not interact either with the electron orbital momentum since the latter has a s-like symmetry. The hole orbital however, is p-like and lattice deformations can thus couple different hole orbital states with each other (cf. ref [19] of the main text). As a result, an acoustic phonon can turn a $J_z=1$ polariton into a $J=2$ dark exciton, but it cannot directly turn into another $J_z=-1$ polariton of opposite spin. Since excitons with $J=2$ have to be an intermediate state in the polariton spin flip, the latter is largely inhibited by the fact that the $J=2$ exciton level lies energetically close to the exciton bright state, and hence due to the magnitude of the Rabi splitting, very far from the polariton ground state.

We should thus observe a vanishing \emph{fast} ASF signal in a cross-polarized detection scheme. However, on the contrary, we observe a signal which is strongly depolarized and isotropic: the latter contributes by $>50\%$ in the wavevector area comprised in between the red and orange dashed lines ($1\mu m^{-1}<k_\parallel<4\mu m^{-1}$) in Fig\ref{figS2} panels a b and c. Two mechanisms are responsible for this depolarization and isotropic character :

\begin{itemize}
\item Semiconductor MCs exhibits significant TE/TM splitting due to the dielectric nature of the layers \cite{langbein2}. Ours is quite large ($2.67\times10^{-14}$ meV.m$^2$) and thus affects significantly the polarization of anti-Stokes polaritons during their lifetime. This interaction is at the basis of effects such as the optical spin Hall effect \cite{kavokin,bramati}, where a coherent polariton field pumped by a linearly polarized laser at $k_\parallel>0$ is scattered by disorder. The TE/TM splitting causes a spin precession along the four diagonal directions with respect to the laser polarization $\phi_k=[\pi/4,3\pi/4,5\pi/4,7\pi/4]$ (yellow dash-dotted lines in Fig.\ref{figS2}), while the laser polarization is conserved along the four directions $\phi_k=[0,\pi/2,\pi,3\pi/2]$ (white dotted lines in Fig.\ref{figS2}). We have checked that this precession indeed occurs as expected in our MC by carrying out a resonant excitation experiment (cf. Fig.\ref{figS2}.e and f). In our cross-polarized detection scheme, the four lobes where spin precession takes place are clearly visible along the four diagonal directions. For anti-Stokes polariton, a very similar behavior is expected, except that since the laser coherence is lost in the scattering process, the spin precession is replaced by a depolarization along the diagonal directions. The linear polarization degree should thus look like the red dotted line in $Fig\ref{figS2}.c$: four lobes of depolarized light along the diagonal directions, and fours lobes of colinearly polarized light along the four other directions. However, the measured degree of polarization $\rho_p(\phi_k)$ is shown as a black solid line in the same plot: it is mostly flat, with no such lobes. We thus understand how depolarization occurs for half of the scattered polaritons but a second mechanism needs to be involved to explain the measured $\rho_p(\phi_k)$.

\item During their lifetime, anti-Stokes polaritons interact strongly with the MC in-plane disorder. In this case, the depolarized polaritons gets scattered over every directions isotropically (Rayleigh scattering), while linearly polarized polaritons gets depolarized in the process, since this scattering mechanism is also affected by the TE/TM splitting. This additional step thus results in a fully depolarized \emph{fast} polariton ASF.

\end{itemize}

This depolarization process of \emph{fast} ASF is a useful advantage for our experiment since, although we detect cross-polarized with respect to the laser, we miss only about one-half of the emitted photons for any of the cooling or heating mechanisms. Moreover, the isotropic character of the emission allows us to extrapolate the total ASF from an incomplete measurement (a measurement along a single slice in momentum space, defined by the monochromator slit).

\subsection{Theoretical model for the \emph{fast} cooling mechanism}

In order to describe the \textit{fast} cooling mechanism theoretically we consider a 2D polariton condensed gas immersed in a 3D phonon bath. To determine the polariton transition rate $W_{k\rightarrow k'}$ from an initial momentum $k$ to a final momentum $k'$ mediated by phonon absorption we use the Fermi golden rule:
\begin{equation}
W_{k\rightarrow k'} = \frac{2 \pi}{\hbar} \sum_{q_z,q_{\|}}
\left|\left\langle k' \right| \left\langle n_{q_{\|},q_z} -1\right| H_{pol-ph}\left|
n_{q_{\|},q_z} \right\rangle \left| k \right\rangle \right|^2
\delta(E_{pol}(k')-E_{pol}(k) -E_{ph}(q_{\|},q_z)) \delta_{k',k+q_{\|}}
\end{equation}
where
\begin{equation}
H_{pol-ph} = \sum_{q'_{z}} \sum_{q'_{\|},\tilde k,\tilde k'}
\mathcal{X}_{\tilde k} \mathcal{X}_{\tilde k'} ~ G(q'_{\|},q'_{z})
\delta_{\tilde k',\tilde k+ q'_{\|}} (c_{q'_{\|},q'_{z}} - c^{\dagger}_{-q'_{\|},q'_z})
b^{\dagger}_{\tilde k'} b_{\tilde k},
\end{equation}
with $\mathcal{X}_{\tilde k}$'s being the Hopfield coefficients, $b^{\dagger}_{k} (b_{k})$ is the creation (destruction) operator of 2D exciton with in-plane wave vector $k$. $c^{\dagger}_{q'_{\|},q'_{z}}(c_{q'_{\|},q'_{z}})$ is the creation (destruction) operator of 3D phonons with wave vector $(q'_{\|},q'_z)$, and
\begin{equation}
G(q'_{\|},q'_{z})=i \sqrt{\frac{\hbar (|q'_{\|}|^2 +{q'_{z}}^2)^{1/2}}{2 \rho V u}}
[D_e I^{\|}_e(|q'_{\|}|)I^{\perp}_e(q'_{z}) -
 D_h I^{\|}_h(|q'_{\|}|)I^{\perp}_h(q'_{z})] ,
\end{equation}
where $u=3.5\times10^3$ m.s$^{-1}$, $\rho=5.65\times10^3$ kg.m$^{-3}$ are the longitudinal sound velocity and the density in ZnSe, and the quantization volume is given by $V$. $D_e=-13.26$ eV and $D_h=-6.56$ eV are the deformation potential for the electron and hole. $I^{\|}_{e(h)}(|q'_{\|}|)$ and $I^{\perp}_{e(h)}(q'_{z})$ are the overlap integrals between the electron (hole) bound in a $1s$ quantum well exciton state and the phonon-modes. As derived in \cite{tassone1,piermarocchi},
\begin{eqnarray}
I^{\|}_{e(h)}(|q'_{\|}|)&=&\left[1+\left(\frac{m_{e(h)}}{2M}|q'_{\|}|a_B\right)^2\right]^{-3/2} \\
I^{\perp}_{e(h)}(q'_z)&=&\text{FT}\{|f_{e(h)}(z)|^2\}
\end{eqnarray}
where $m_e=0.18$ and $m_h=1.26$ are the electron/hole mass in electron mass units, $M=m_e+m_h$ is the exciton mass and $a_B=4.1$ nm is the bulk ZnSe exciton Bohr radius. $\text{FT\{\}}$ is the one-dimensional spatial Fourier transform, and $f_{e(h)}$ is the electron (hole) quantum well wavefunction along the confinement axis $z$. The latter is calculated according to our actual quantum wells parameters: $L_z=8$ nm thickness and $310$ meV confinement energy.
Finally,
\begin{equation}
W^{f}_{k\rightarrow k'} =  \frac{2 \pi}{\hbar} \sum_{q_z,q_{\|}}
n({q_{\|},q_z}) |\mathcal{X}_{k}|^2 |\mathcal{X}_{k'}|^2
|G(q_{\|},q_{z})|^2 \delta_{k',k+ q_{\|}}
\delta[E^{f}_{pol}(k')-E^{f}_{pol}(k) -E_{ph}(q_{\|},q_z)],
\label{wkk:free}
\end{equation}
where $E^{f}_{pol}(k)$ is the dispersion of noninteracting polaritons, the phonon occupation is given by the Bose-Einstein distribution $n({q_{\|},q_z})=(\exp[E_{ph}(q_{\|},q_z)/(k_B T)] - 1)^{-1}$, with $E_{ph}=\hbar u \sqrt{q_{\|}^2+q_z^2}$ the dispersion of acoustic phonons. In order to calculate the ASF spectrum, $k=0$ is taken as the initial state. We Include a finite linewidth for the polariton dispersion so that Eq. (\ref{wkk:free}) becomes
\begin{eqnarray}
W^{f}_{0 \rightarrow k'} &=&  \frac{2 \pi}{\hbar \pi} \sum_{q_z,q_{\|}}
n({q_{\|},q_z}) |\mathcal{X}_{0}|^2 |\mathcal{X}_{k'}|^2
|G(q_{\|},q_{z})|^2 \delta_{k',0+ q_{\|}}
\frac{\gamma}{[E^{f}_{pol}(k')-E_{ph}(q_{\|},q_z)]^2+\gamma^2}.
\end{eqnarray}

The ZnSe material parameters have been taken from refs. \cite{rudin,cardona}.

\subsection{Orders of magnitude of the pump polariton non-radiative recombination rate}

\begin{figure}[t]
\includegraphics[width=5cm]{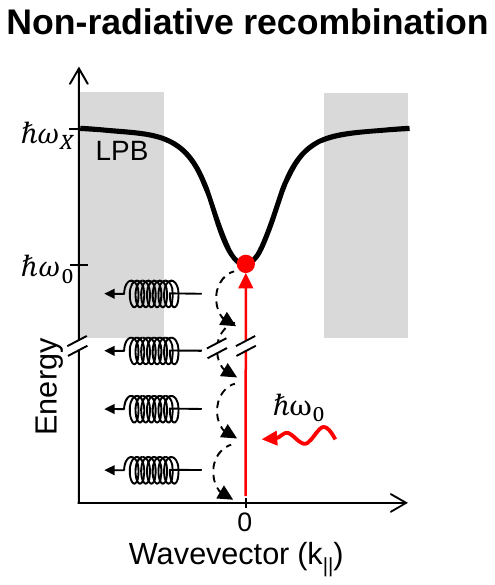}
\caption{(Color online) Schematic representation of non-radiative recombination of pumped polaritons of energy $\hbar\omega_0$. The intermediate states of the cascade are point-like defects in the crystalline structure. The multi-phonon cascade generates a total thermal energy of $\hbar\omega_0$ in the phonon bath}
\label{figS3}
\end{figure}

Non-radiative recombination (NRR) is a process where an excitation (a pump polariton in our case) relaxes its total energy in the form of a cascade of phonons, involving localized defect states lying in the material bandgap (cf. Fig.\ref{figS3}). It is thus a source of heat. Owing to their largely dominant population, NRR of pump polaritons is the largest potential source of NRR in our microcavity. To our knowledge, NRR of exciton-polaritons has never been measured so far. Indeed, such a measurement is quite challenging as the NRR lifetime is around 6 orders of magnitude longer than the polariton lifetime in the MC.

In order to estimate $P_\text{NRR}$, the contribution of pump polariton NRR to the heating power, we make the rough approximation the polariton NRR rate $\gamma_{nr}$ is that of exciton multiplied by the excitonic fraction $X^2$. Then, Within a simple rate equation model, $P_\text{NRR}=P_\text{las}X^2\hbar\omega_0\gamma_\text{nr}/\gamma$, where $\gamma$ is the polariton radiative rate. In this approximation, we find that
$P_\text{NRR}$ is equal to $P_\text{fr}^\text{max}$ for $\gamma_\text{nr}^{-1}\simeq 1500$ ns.

To get an order of magnitude to compare with, the above estimation can be compared with the excitonic NRR rate $\gamma_\text{nr,X}^{-1}=350$ns found for excitons in ref. \cite{nr2}, in Zn$_{0.50}$Cd$_{0.50}$Se/Zn$_{0.21}$Cd$_{0.19}$Mg$_{0.60}$Se, n-doped quantum wells (QWs) with a thickness of $d=4.5$nm, measured at T=$50K$. This lower than $\gamma_\text{nr}^{-1}$, however, in our microcavity, we are dealing with a different material and structure, and with polaritons instead and excitons. Thus, several qualitative arguments advocate for a smaller $\gamma_\text{nr}$ by at least a few orders of magnitude:

\begin{itemize}

\item
The first set of arguments concerns the structural properties: (i) the material of our QWs is binary ZnSe, so that the defect density within the QW is thus much lower than a for a ternary material. (ii) Our QWs have a thickness of $d=8$nm and are thus twice thicker. Since most defect leading to NRR are formed at the interface between the QW and the barrier, an approximately twice lower defect density is expected. (iii) Some of us have carried out a TEM investigations of our MC, including the QWs. The defect density which is found is indeed very low as and sets the state of the art for epitaxial II-VI materials \cite{TEM}. (iv) The electron donor responsible for the n-doping provides non-negligible density of defect state, possibly contributing to NRR.

\item
The main argument relies on the polariton mass: in the comparison above, it is assumed that the coupling between a polariton and a defect state is only reduced by a factor $X^2$ as compared to an exciton. However, considering the polariton mass typically four orders of magnitude lighter than the exciton one, and according to the Fermi golden rule, its capture by a point-like defect of a size comparable with the lattice parameters must be reduced by the same order of magnitude.

\end{itemize}

With this set of arguments in mind, we are quite confident that the actual NRR lifetime $\gamma_\text{nr}^{-1}$ in our MC exceeds significantly $1500$ ns and can thus be reasonably neglected.

\end{widetext}

\end{document}